\documentstyle[psfig,graphics]{mn2e}
\title [Mass loss and orbital period decrease in CABs]
{Mass loss and orbital period decrease
in detached chromospherically active binaries}

\author[Demircan et al.]
       {O. Demircan,$^1 \thanks{E-mail: demircan@comu.edu.tr}$
        Z. Eker$^{1}$, Y.~Karata\c{s}$^2$ and S. Bilir$^{2}$\\ 
  $^1$\c{C}anakkale Onsekiz Mart University, Ulup\i nar Astrophysical Observatory, 
      17100 \c{C}anakkale, Turkey\\
  $^2$Istanbul University Science Faculty, Department of Astronomy and Space
      Sciences, 34119, University-Istanbul, Turkey\\}

\date{Accepted 2005 month day.
      Received year month day;
      }
\pagerange{\pageref{firstpage}--\pageref{lastpage}}
\pubyear{2005}

\begin{document}
\maketitle
\label{firstpage}
\begin{abstract}
The secular evolution of the orbital angular momentum (OAM), the systemic 
mass $(M=M_{1}+M_{2})$ and the orbital period of 114 chromospherically active 
binaries (CABs) were investigated after determining the kinematical ages of the 
sub-samples which were set according to OAM bins. OAMs, systemic masses and 
orbital periods were shown to be decreasing by the kinematical ages. The first 
order decreasing rates of OAM, systemic mass and orbital period have been 
determined as $\dot J = 3.48 \times 10^{-10}~yr^{-1}$ per systemic OAM, 
$\dot M = 1.30 \times 10^{-10}~yr^{-1}$ per systemic mass and  
$\dot P = 3.96\times 10^{-10}~yr^{-1}$ per orbital period respectively from the 
kinematical ages. The ratio of $d \log J/ d \log M  = 2.68$, which were derived 
from the kinematics of the present sample, implies that there must be a mechanism 
which amplifies the angular momentum loss $\bar A = 2.68$ times in comparison to 
isotropic angular momentum loss of hypothetical isotropic wind from the 
components. It has been shown that simple isotropic mass loss from the surface 
of a component or both components would increase the orbital period.   
\end{abstract}
\begin{keywords}
stars: mass-loss, stars: evolution, stars: binaries spectroscopic 
\end{keywords}

\section{Introduction}
The observational evidence of decaying rotation rate for stars with spectral 
types later than F in the stellar evolution was well documented by Skumanich\ 
(1972) by studying projected equatorial speeds $(v \sin i)$ of the late-type 
stars in open clusters of different ages. Such a decay process in stellar 
rotation is explained in the terms of angular momentum loss (AML) through 
magnetically driven stellar winds, also called the magnetic braking (cf. 
Schatzman\ 1959, Kraft\ 1967 and Mestel\ 1968). For the tidally locked binaries 
with late-type components, such an AML is known to be provided by the reservoir 
of the orbital angular momentum (OAM). Therefore, the AML from the components of 
spin-orbit coupled binaries causes the orbit to shrink. Spin-orbit coupling and 
a shrinking orbit, then, imposes spin-up the rotation of the components, which 
is different from single stars slowing down rotation. This mechanism is 
considered as the main way to form W UMa-type contact binaries from systems 
initially detached (cf. Huang\ 1966, Okamoto \& Sato\ 1970, van't Veer\ 1979, 
Vilhu \& Rahunen\ 1980, Mestel\ 1984, Guinan \& Bradstreet\ 1988, Maceroni \& 
van't Veer\ 1991, Stepien\ 1995, Demircan\ 1999).

The period evolution and the time scale of forming the contact binaries from the 
detached progenitors were estimated differently among the various authors. In the 
work of Guinan \& Bradstreet\ (1988), the AML of a component star was computed 
directly from Skumanich's law ($V_{rot} \sim t^{1/2}$), which is derived from the 
relatively slow rotating stars ($V_{rot}\leq 17$ km s$^{-1}$), and then AML from 
the two components made equal to the orbital AM change. Orbits evolve initially 
almost with constant periods until the very end where the orbits shrink sharply
to form contact binaries. For the braking law of binary orbits, van't Veer \& 
Maceroni\ (1988, 1989) gave a period evolution function which was predicted from 
the initial (Abt \& Levy\ 1976, Abt\ 1983) and the present day (Farinella et al.\ 
1979) period distributions of G-type binaries corrected for selection and 
detectability effects. Stepien\ (1995) derived a formula with no free parameters 
for the AML via a magnetized wind and calibrated it from the spin down of single 
stars. He concludes that usually the orbital periods which are less than 5 days 
can form contact systems within their main-sequence lifetime.

What is common among those works is the spin-orbit coupling set by synchronization. 
In other words, the above mechanisms do not work for asynchronous binaries as van't 
Veer \& Maceroni\ (1988) stated that the wider systems without spin-orbit coupling 
will not change the orbital period. Therefore, Demircan\ (1999) studied the orbital 
AM distribution of 40 well-known CABs only with $P_{orb}<10$ days, and found an 
observational estimate of the rate of orbital AML and the braking law from the upper 
boundary of the orbital AM distribution.

Usage of small number statistics and saturation in activity-rotation relation 
may be responsible for the weakness among those studies. Perhaps, the upper
boundary does not represent all systems, and the mass loss and the AML may be 
age-dependent quantities. In order to better understand the orbital period evolution
of detached active binaries, systems with different ages should be considered. 
Although the ages of binaries can be estimated only by detailed evolutionary isochrones,
the sub-group ages can be found kinematically. Recently, after having more accurate 
data on a large sample of detached CABs, Karata\c{s} et al.\ (2004) became successful 
in breaking up the sample into kinematically distinct sub-samples. Karata\c{s} et al.\ 
(2004) initially divided the whole sample of 237 systems into two groups: the possible 
moving group (MG) members (95) and the older field binaries (142). A comparison of 
the total mass, the orbital period, the mass ratio and the orbital eccentricity of 
these two groups revealed clear observational evidence that detached CABs lose mass 
and angular momentum so that their orbits circularize and shrink. Related orbital 
period decreases, unfortunately, are not detectable on commonly used O--C diagrams 
formed by the eclipse minimum times.
This is due to: 1) very short time span covered by existing O--C data (at most 100 
years) in comparison with durations implied by the predicted kinematical ages which 
are at the order of $10^9$ years; 2) large scatter of unevenly distributed data 
(especially visual and photographic observations) in the present O--C diagrams; 3) 
existence of complicated larger amplitude short time scale fluctuations caused by 
many different effects (cf Kreiner, Kim \& Nha\ 2001; Demircan\ 2000, 2002). Being 
independent of the physical cause, the O--C diagrams are commonly used in the 
study of orbital period changes of binaries in general. At present, mean minimum 
time deviations as small as 10 seconds for binaries with sharp eclipses are detectable. 
Nevertheless, the minimum detectable period variation on O--C diagrams depends on 
the time span of observations over the actual variations.

This classical approach with O--C diagrams, thus, are not suitable to obtain slower  
secular orbital decreases implied by the wind driven mass loss of our sample of 
detached CABs. Consequently, our aim in the present work is to further investigate the 
mass-loss, the angular momentum loss, orbital period decrease and then determine the 
rates of change of those parameters statistically from the new and more accurate data 
(absolute dimensions and kinematical data) of Karata\c{s} et al.\ (2004). The present 
approach, apparently, is more advantageous in detecting changes at the order of 
evolutionary time scales, where O--C diagrams becomes insufficient, and also more 
practical to give no distinction weather the involved binaries are eclipsing or not.

\section{data}
Among the total 237 CABs studied kinematically by Karata\c{s} et al.\ (2004), 
119 systems with complete basic data (orbital and physical), which allows 
computation of orbital angular momentum $J$, were selected for this study.
Although, contact (W UMa) and semi-contact ($\beta$ Lyrea and classical Algols) 
systems were intentionally excluded from the catalog of chromospherically active 
binary stars (CAB, Strassmeier et al.\ 1988, 1993), thus, CABs are known to be 
detached systems. Nevertheless, the catalog still contains small number of systems 
like $\epsilon$ UMi (de Medeiros \& Udry\ 1999), RT Lac (Heunemoerder \& Barden\ 
1986; Popper\ 1991), RV Lib (Popper\ 1991), BH CVn (Eker\ 1987), RZ Cnc (Olson\ 1989), 
AP Psc (Eaton \& Barden\ 1988) and AR Mon (Williamon et al.\ 2005) with a component 
filling or about to fill its Roche lobe that mass transfer possibly exist. Since 
we are interested in only detached active binaries, five of such systems (BH CVn,
RT Lac, RV Lib, AR Mon and $\epsilon$ UMi), were further discarded from our final 
list that our sample reduced to 114 systems.   
  
62 out of 114 are in the sub-sample (142 binaries) which were called the field 
stars. Karata\c{s} et al.\ (2004) assigned an age 3.86 Gyr to this group from the 
galactic space velocity components. This group is a mixture of young and 
old stars together. Therefore, the assigned age does not represent each system 
individually. It is an average age suggested by their kinematics. The 62 field 
stars and their data are displayed in Table 1. Columns are order number, name, HD 
number, orbital period, rotational period of the most active star, total mass  
($M = M_{1} + M_{2}$), mass ratio ($q = M_{2}/ M_{1}<1$), logarithm of the orbital 
angular momentum and assigned kinematical age. How these ages were assigned will be 
explained in Section 3.2.

Those field binaries have been further divided in five groups according to OAM but 
displayed as sorted on orbital periods. The groups are shown as separated by a blank 
row in Table 1.     

\begin{table*}
\begin{minipage}{180mm}
\center
\caption{Physical parameters of the field CABs.}
{\scriptsize
\begin{tabular}{lllrrcccc}
\hline
        No &       Name &         HD &   $P_{orb}$  &  $P_{rot}$\footnote{from Strassmeier et al.\ (1993)}
           & $M _{tot}$   & $q$ &   $\log J$ &  Age \\
           &            &           &      (days)   &    (days)& $(M _{\odot})$ &     &  (cgs)     & (Gyr)\\
\hline
        1 &     XY UMa &     237786 &      0.479 &            &       3.58 &      0.954 &      51.76 &       9.16 \\
        2 &     BI Cet &       8358 &      0.516 &      0.520 &       1.84 &      0.957 &      51.84 &       9.16 \\
        3 &     SV Cam &      44982 &      0.593 &      0.593 &       1.37 &      0.957 &      51.81 &       9.16 \\
        4 &     WY Cnc &            &      0.829 &      0.829 &       2.82 &      0.958 &      51.75 &       9.16 \\
        5 &     CM Dra &            &      1.268 &      1.268 &       2.75 &      0.833 &      50.94 &       9.16 \\
        6 &     CC Eri &      16157 &      1.562 &      1.561 &       1.44 &      0.675 &      51.42 &       9.16 \\
        7 &     EI Eri &      26337 &      1.947 &      1.945 &       1.18 &      0.269 &      51.63 &       9.16 \\
        8 &     BK Psc &            &      2.166 &            &       1.50 &      0.974 &      51.59 &       9.16 \\
        9 &  V1396 Cyg &            &      3.276 &            &       0.66 &      0.696 &      51.35 &       9.16 \\
       10 &     HZ Aqr &            &      3.757 &            &       1.23 &      0.804 &      51.83 &       9.16 \\
       11 &     UZ Lib &            &      4.768 &      4.736 &       1.54 &      0.921 &      51.84 &       9.16 \\
       12 &     BY Dra &     234677 &      5.975 &            &       5.09 &      0.951 &      51.70 &       9.16 \\
       13 &     LR Hya &      91816 &      6.866 &      3.145 &       2.91 &      0.942 &      51.83 &       9.16 \\
       14 &     II Peg &     224085 &      6.724 &      6.718 &       1.20 &      0.500 &      51.85 &       9.16 \\
       15 &     OU Gem &      45088 &      6.992 &      7.360 &       1.43 &      0.700 &      51.89 &       9.16 \\
    &  &  &  &  & &  &  &  \\
       16 &     UV Leo &      92109 &      0.600 &            &       3.06 &      0.951 &      51.99 &       6.57 \\
       17 &     BH Vir &     121909 &      0.817 &      0.817 &       3.53 &      0.604 &      51.97 &       6.57 \\
       18 &   V837 Tau &      22403 &      1.930 &      1.890 &       2.12 &      0.909 &      51.94 &       6.57 \\
       19 &     AR Lac &     210334 &      1.983 &      1.983 &       2.39 &      0.897 &      52.22 &       6.57 \\
       20 &     CF Tuc &       5303 &      2.798 &      2.798 &       1.75 &      0.865 &      52.23 &       6.57 \\
       21 &     AD Cap &     206046 &      2.960 &      2.960 &       1.62 &      0.526 &      51.96 &       6.57 \\
       22 &     AS Dra &     107760 &      5.413 &      5.413 &       2.54 &      0.906 &      51.99 &       6.57 \\
       23 &  V1423 Aql &     191262 &      5.434 &      5.530 &       1.85 &      0.682 &      52.25 &       6.57 \\
       24 &     KT Peg &     222317 &      6.202 &      6.092 &       1.54 &      0.671 &      52.05 &       6.57 \\
       24 &     UV CrB &     136901 &     18.667 &            &       4.10 &      0.464 &      52.04 &       6.57 \\
    &  &  &  &  & &  &  &  \\
       26 &   V711 Tau &      22468 &      2.838 &      2.841 &       1.85 &      0.993 &      52.31 &       4.07 \\
       27 &     PW Her &            &      2.881 &      2.881 &       2.50 &      0.786 &      52.34 &       4.07 \\
       28 &     TY Pyx &      77137 &      3.199 &      3.320 &       2.22 &      0.947 &      52.30 &       4.07 \\
       29 & SAO 240653 &     114630 &      4.233 &            &       3.25 &      0.979 &      52.26 &       4.07 \\
       30 &     UX Ari &      21242 &      6.438 &      6.438 &       2.19 &      0.975 &      52.28 &       4.07 \\
       31 &     SS Boo &            &      7.606 &      7.606 &       1.04 &      0.951 &      52.26 &       4.07 \\
       32 &     EZ Peg &            &     11.660 &     11.663 &       1.86 &      0.991 &      52.30 &       4.07 \\
       33 &     AR Psc &       8357 &     14.302 &     12.245 &       1.63 &      0.567 &      52.39 &       4.07 \\
       34 &   V350 Lac &     213389 &     17.753 &            &       1.80 &      0.818 &      52.33 &       4.07 \\
       35 &     FG UMa &      89546 &     21.360 &            &       2.83 &      0.991 &      52.37 &       4.07 \\
       36 &     XX Tri &      12545 &     23.969 &     24.300 &       1.67 &      0.964 &      52.30 &       4.07 \\
    &  &  &  &  & &  &  &  \\
       37 &     SZ Psc &     219113 &      3.965 &      3.955 &       2.86 &      0.766 &      52.44 &       2.95 \\
       38 &      Z Her &     163930 &      3.993 &      3.962 &       1.99 &      0.881 &      52.45 &       2.95 \\
       39 &     RS UMi &            &      6.169 &            &       3.09 &      0.818 &      52.41 &       2.95 \\
       40 &     MM Her &     341475 &      7.960 &      7.936 &       1.70 &      0.546 &      52.45 &       2.95 \\
       41 &     FF Aqr &            &      9.208 &      9.208 &       3.10 &      0.240 &      52.43 &       2.95 \\
       42 &     42 Cap &     206301 &     13.174 &            &       2.37 &      0.727 &      52.48 &       2.95 \\
       43 &     IS Vir &     113816 &     23.655 &            &       0.62 &      0.938 &      52.44 &       2.95 \\
       44 &     AI Phe &       6980 &     24.590 &            &       1.19 &      0.924 &      52.60 &       2.95 \\
       45 &     IM Peg &     216489 &     24.649 &     24.390 &       2.30 &      0.533 &      52.52 &       2.95 \\
       46 &     TW Lep &      37847 &     28.344 &     28.220 &       2.78 &      0.704 &      52.47 &       2.95 \\
       47 &  V4200 Ser &     188088 &     46.817 &     16.500 &       4.93 &      0.904 &      52.43 &       2.95 \\
       48 &     EL Eri &      19754 &     48.263 &     47.960 &       0.50 &      0.917 &      52.43 &       2.95 \\
       49 &    AY Cet  &       7672 &     56.824 &     77.220 &       1.75 &      0.765 &      52.60 &       2.95 \\
       50 &  V1197 Ori &      38099 &    143.040 &            &       1.29 &      0.949 &      52.55 &       2.95 \\
    &  &  &  &  & &  &  &  \\
       51 &     RU Cnc &            &     10.173 &     10.135 &       2.23 &      0.855 &      52.61 &       2.55 \\
       52 &     CQ Aur &     250810 &     10.623 &     10.560 &       1.83 &      0.564 &      52.79 &       2.55 \\
       53 &     TZ Tri &      13480 &     14.729 &     14.729 &       2.73 &      0.365 &      53.06 &       2.55 \\
       54 &$\zeta$ And &       4502 &     17.769 &            &       2.14 &      0.739 &      52.65 &       2.55 \\
       55 &     BL CVn &     115781 &     18.692 &     18.692 &       0.57 &      0.979 &      52.63 &       2.55 \\
       56 &     CS Cet &       6628 &     27.322 &            &       2.15 &      0.957 &      52.71 &       2.55 \\
       57 &   V792 Her &     155638 &     27.538 &     27.070 &       4.50 &      0.800 &      52.74 &       2.55 \\
       59 &  V1762 Cyg &     179094 &     28.590 &     28.590 &       7.73 &      0.600 &      52.72 &       2.55 \\
       59 &   V965 Sco &     158393 &     30.969 &     30.960 &       3.95 &      0.879 &      52.88 &       2.55 \\
       60 &     RZ Eri &      30050 &     39.282 &     31.400 &       1.10 &      0.970 &      52.89 &       2.55 \\
       61 &     DK Dra &     106677 &     64.474 &     63.750 &       2.55 &      0.930 &      53.00 &       2.55 \\
       62 &  BD+44 801 &      23838 &    962.800 &            &       0.68 &      0.545 &      53.22 &       2.55 \\
\hline
\end{tabular} 
}
\end{minipage}
\end{table*}

\begin{table*}
\begin{minipage}{150mm}
\center
\caption{Physical parameters of the MG CABs.}
{\scriptsize
\begin{tabular}{lllrrccclll}
\hline
 No &  Name &  HD & $P_{orb}$ & $P_{rot}$\footnote{from Strassmeier et al.\ (1993)}
           & $M_{tot}$     &  $q$ &  $\log J$ & MG &   OC &   Age \\
           &       &     & (days)    &  (days)  & $(M _{\odot})$&    & (cgs)  &    &      &  (Gyr) \\
\hline
1 & V471 Tau    &         &  0.521 &  0.597 & 1.50 & 0.974 & 51.69 & Hya     &     Hya  &      0.6 \\
2 & RT And      &         &  0.629 &  0.629 & 2.14 & 0.739 & 51.97 & UMa     &          &      0.3 \\
3 & CG Cyg      &         &  0.631 &  0.631 & 1.75 & 0.865 & 51.83 & UMa     &          &      0.3 \\
4 & ER Vul      & 200391  &  0.698 &  0.694 & 2.15 & 0.957 & 51.99 & IC      &          &    0.055 \\
5 & YY Gem      & 60179   &  0.814 &  0.814 & 1.19 & 0.924 & 51.59 & Cas     &          &     0.20 \\
6 & UV Psc      & 7700    &  0.861 &  0.861 & 1.75 & 0.765 & 51.87 & IC      &          &    0.055 \\
7 & V1430 Aql   &         &  0.874 &        & 1.84 & 0.957 & 51.91 & LA      &          &     0.15 \\  
8 & V772 Her    & 165590  &  0.880 &  0.878 & 1.63 & 0.567 & 51.79 & LA      &          &     0.15 \\ 
9 & IL Com      & 108102  &  0.962 &  0.820 & 1.67 & 0.964 & 51.86 & Cas     & Com Ber  &     0.20 \\
10 & $\delta$ Cap & 207098&  1.023 &        & 2.73 & 0.365 & 52.12 & LA      &          &     0.15 \\ 
11 & DH Leo      & 86590  &  1.070 &  1.066 & 1.44 & 0.675 & 51.75 & Hya, IC &          &      0.6 \\
12 & Gl 841A     &        &  1.123 &        & 0.50 & 0.917 & 51.01 &  LA     &          &     0.15 \\  
13 & TZ CrB      & 146361 &  1.140 &  1.169 & 2.19 & 0.975 & 52.08 & LA      &          &     0.15 \\ 
14 & BD+23 2297  & 95559  &  1.526 &  1.526 & 1.85 & 0.993 & 52.00 & Hya, IC &          &      0.6 \\  
15 & V824 Ara    & 155555 &  1.682 &  1.682 & 2.12 & 0.909 & 52.11 & LA      &          &     0.15 \\  
16 & 13 Cet      & 3196   &  2.082 &        & 0.68 & 0.545 & 51.28 & Hya     &          &      0.6 \\
17 & V478 Lyr    & 178450 &  2.131 &  2.131 & 1.18 & 0.269 & 51.55 & IC      &          &    0.055 \\  
18 & FF And      &        &  2.170 &  2.170 & 1.10 & 0.970 & 51.67 & IC      &          &    0.055 \\ 
19 & KZ And      & 218738 &  3.033 &  3.030 & 1.29 & 0.949 & 51.84 & Cas, IC &          &      0.2 \\ 
20 & BD+39 4529  & 203454 &  3.243 &        & 1.83 & 0.564 & 52.06 & UMa     &          &      0.3 \\
21 & V835 Her    & 163621 &  3.304 &  3.350 & 1.43 & 0.700 & 51.91 & LA      &          &     0.15 \\ 
22 & HZ Com      &        &  3.558 &        & 1.37 & 0.957 & 51.90 & Cas     & Com Ber  &      0.2 \\
23 & GK Hya      &        &  3.587 &  3.587 & 2.56 & 0.910 & 52.36 & Hya     &          &      0.6 \\
24 & UX Com      &        &  3.642 &  3.642 & 2.23 & 0.855 & 52.26 & Hya     &          &      0.6 \\
25 & BU 163      & 202908 &  3.966 &        & 2.22 & 0.947 & 52.27 & UMa     &          &      0.3 \\
26 & RS CVn      & 114519 &  4.798 &  4.791 & 2.82 & 0.958 & 52.47 & IC      &          &    0.055 \\  
27 & SS Cam      &        &  4.824 &  4.823 & 3.58 & 0.954 & 52.64 & LA      &          &     0.15 \\ 
28 & RT CrB      & 139588 &  5.117 &  5.117 & 2.83 & 0.991 & 52.48 & UMa     &          &      0.3 \\
29 & VV Mon      &        &  6.051 &  6.051 & 2.91 & 0.942 & 52.53 & IC      &          &    0.055 \\ 
30 & RW UMa      &        &  7.328 &  7.328 & 3.06 & 0.951 & 52.59 & IC      &          &    0.055 \\ 
31 & LX Per      &        &  8.038 &  7.905 & 2.55 & 0.930 & 52.47 & Hya     & $\alpha$ Per&   0.6 \\
32 & AW Her      & 348635 &  8.801 &        & 2.54 & 0.906 & 52.48 & Hya     &          &      0.6 \\
33 & V1285 Aql   &        & 10.319 &  2.900 & 0.62 & 0.938 & 51.48 & Cas     &        &        0.2 \\
34 & AE Lyn      & 65626  & 11.068 & 10.163 & 3.25 & 0.979 & 52.69 & IC      &        &      0.055 \\ 
35 & V829 Cen    & 101309 & 11.710 & 11.660 & 0.57 & 0.979 & 51.45 & Cas     &        &        0.2 \\
36 & V808 Tau    & 283882 & 11.929 &  6.820 & 1.58 & 0.950 & 52.18 & Hya     &   Hya  &        0.6 \\
37 & IL Hya      & 81410  & 12.905 & 12.890 & 3.53 & 0.604 & 52.75 & UMa     &        &        0.3 \\
38 & V1379 Aql   & 185510 & 20.661 & 25.640 & 3.05 & 0.129 & 52.34 & LA      &        &       0.15 \\  
39 & ADS11060C   & 165590C& 25.763 &  9.000 & 1.04 & 0.951 & 51.99 & LA      &        &       0.15 \\  
40 & BD+64 487   & 30957  & 44.396 &        & 1.54 & 0.921 & 52.35 & Cas, IC &        &        0.2 \\ 
41 & KX Peg      & 212280 & 45.284 & 29.060 & 3.09 & 0.818 & 52.86 & UMa     &        &        0.3 \\
42 & BD+44 2760  & 161570 & 45.623 &        & 2.75 & 0.833 & 52.77 & IC      &        &      0.055 \\ 
43 & GT Mus      & 101379 & 61.360 & 56.030 & 4.50 & 0.800 & 53.17 & LA      &        &       0.15 \\  
44 & DQ Leo      & 102509 & 71.690 & 55.000 & 3.95 & 0.879 & 53.10 & Hya     &        &        0.6 \\
45 & BD+17 703   & 27149  & 75.648 &        & 1.99 & 0.881 & 52.62 & Hya     &   Hya  &        0.6 \\
46 & BM Cam      & 32357  & 80.898 & 85.000 & 1.70 & 0.546 & 52.47 & LA      &        &       0.15 \\ 
47 & 5 Cet       & 352    & 96.400 & 96.320 & 2.50 & 0.786 & 52.81 & Cas     &        &        0.2 \\
48 & $\alpha$ Aur & 34029 &104.023 & 80.000 & 5.09 & 0.951 & 53.34 & Hya     &        &        0.6 \\
49 & V1817 Cyg   & 184398 &108.854 &108.854 & 7.73 & 0.600 & 53.62 & UMa     &        &        0.3 \\
50 & $\eta$ And  & 5516   &115.720 &        & 4.93 & 0.904 & 53.33 & UMa     &        &        0.3 \\
51 & SAO 23511   & 57853  &122.169 &        & 1.85 & 0.682 & 52.62 & IC      &        &      0.055 \\ 
52 & V819 Her    & 157482 &2018.000& 81.900 & 2.78 & 0.704 & 53.32 & UMa     &        &        0.3 \\
\hline
\end{tabular} 
}
\end{minipage}
\end{table*}

The rest, 52 systems out of 114, are in the sub-sample which were called MG. 
Kinematical criteria originally defined by Eggen\ (1958a, b, 1989, 1995), 
for determining possible members of the best-documented moving groups, are 
summarized by Montes et al.\ (2001a, b).  One may see Karata\c{s} et al.\ (2004) 
for details as to how MG systems were selected from the common CABs.
However, the basic idea is that a test star's space velocity vector 
must be equal and parallel, or at least with deviations smaller than 
pre-determined limits, to the space velocity vector of a moving group. 
The ages of moving groups are known as open cluster ages by the turn-off 
point from the main sequence. Consequently, a pre-determined age of 
a moving group can be assigned to all binaries which are found to be 
possible members according to their space velocity vectors. Therefore, 
unlike the field stars with various ages, the MG stars are homogeneous
with a single age corresponding to each MG. Among the five MG 
considered by Karata\c{s} et al.\ (2004), the Hyades Supercluster is 
the oldest one with 0.6 Gyr age (see Table 3 of Karata\c{s} et al.\ 2004).
The 52 systems and their assigned ages are listed in Table 2 in order 
of orbital period length. The columns are same as Table 1 but two more 
columns, name of the MG and associated open cluster, were inserted before the 
ages according to identified MG.

\section{OAM evolution among the detached CABs}
\subsection{Basics of mass loss and OAM change}

Assuming component masses as points, the OAM of a binary is given by the 
well known relation

\begin{eqnarray}
J= {M_{1}M_{2}\overwithdelims()M_{1} + M_{2}}{a^2 \Omega} 
= {q \overwithdelims()(1 + q)^{2}}{M a^2 \Omega},
\end{eqnarray}
where
\begin{eqnarray}
{M_{1}M_{2}\overwithdelims()M_{1} + M_{2}}{a^2} 
= {q \overwithdelims()(1 + q)^{2}}{M a^2}= I
\end{eqnarray}
is the moment of inertia and $\Omega = 2\pi /P$ is the angular velocity of the 
system. $P$ is the orbital period, $M = M_{1} + M_{2}$ is the total mass, 
$q = M_{2}/ M_{1}<1$ is the mass ratio of the components and $a$ is the semi 
major axis of the binary. Therefore, an isotropic mass loss from the surfaces of 
components will produce an OAM loss as

\begin{eqnarray}
dJ= {q \overwithdelims()(1 + q)^{2}}{a^2 \Omega dM},
\end{eqnarray}
which might be due to isotropic stellar winds from one or both components under 
the condition that there must be no interaction between the winds nor between 
the winds and the components. Binaries are dynamical systems obeying Kepler's third 
law. Therefore, the orbit re-arranges itself. Thus, the basic parameters ($a$, $M$ 
and $P$) must all change according to 

\begin{eqnarray}
3{da \over a}+ 2{d\Omega \over \Omega} = {dM \over M}.
\end{eqnarray} 
Consequently, the OAM change must provide their changes as 

\begin{eqnarray}
{dJ \over J} = {dM \over M} + 2{da \over a}+ {d\Omega \over\Omega}  
\end{eqnarray} 
according to (1), where the contribution of mass ratio change 
($dq$) is neglected in the first approximation. If ${da \over a}$ 
is eliminated between (4) and (5), then

\begin{eqnarray}
{dJ \over J} = {5\over3}{dM \over M}-{1\over3}{d\Omega \over\ \Omega}
={5\over3}{dM \over M}+ {1\over3}{dP \over P} 
\end{eqnarray} 
is obtained. But if  $d\Omega \over\ \Omega$ is eliminated between (4) and (5),

\begin{eqnarray}
{dJ \over J} = {3\over2}{dM \over M} + {1\over2}{da \over a}
\end{eqnarray} 
is obtained. In another words, OAM change provides two independent equations 
which tell us how OAM is shared among the two parameters, $M$ and $P$ in one case, 
$M$ and $a$ in the other case, where the change in the missing parameter is provided 
by Kepler's third law (eq. 4).

According to (1) and (3), the isotropic mass loss from surfaces implies

\begin{eqnarray}
{dJ \over J} = {dM \over M}, 
\end{eqnarray} 
which means the change in OAM has only one source: the mass loss, which were 
assumed to be isotropic. In order to see how the orbit reacts to this OAM change, 
it is plugged into eq. (6). Then, it gives us

\begin{eqnarray}
{dM \over M} = -{1\over 2}{dP \over P}, 
\end{eqnarray} 
which means the orbital period must increase since negative $dM$ makes $dP$ positive.
But, if it is plugged into eq. (7), it gives us

\begin{eqnarray}
{dM \over M} = -{da \over a}, 
\end{eqnarray} 
which means the semi-major axis of the binary must increase similarly 
because negative $dM$ makes $da$ positive. A similar conclusion is given by 
Pringle\ (1985) who assumed isotropic non-interacting mass loss from one 
component only. Here, the problem is generalized. It does not matter whether 
one or both components lose mass, the result is the same; simple mass loss 
from the surfaces of components will cause the orbital period to increase rather
than to decrease. Despite this, the system loses OAM. If a binary does not lose 
mass, the total angular momentum (orbital plus spin angular momentums) of the 
system is conserved. Therefore, for a system to lose angular momentum, mass 
loss is inevitable. However, eq. (8) tells us that ${dJ \over J} = {dM \over M}$, 
that is, each particle leaving the system carries away an OAM which is equal to 
mean orbital angular momentum per mass in the system. There can be various 
mechanisms which can amplify angular momentum loss per particle leaving the 
system with respect to mean OAM per mass in the system. Consequently we define 
an amplification parameter $\bar A$ as

\begin{eqnarray}
{\bar A} = {{dJ \over dM} \over {J \over M}} 
\Rightarrow {dJ \over J}={\bar A}{dM \over M},
\end{eqnarray}
where ${\bar A}$ is the ratio between angular momentum loss per mass and mean 
orbital angular momentum per mass in the system. Notice that ${\bar A}=1$ for 
isotropic mass loss from the surface of one or both components; then equation (11) 
reduces to be eq. (8). Inserting (11) into (6), we can write

\begin{eqnarray}
{(3\bar A -5)}{dM \over M}  = {dP \over P}. 
\end{eqnarray}
From above, we can set up a condition  $\bar A > 5/3$  for a system to decrease 
its orbital period if orbital period decrease is due to OAM loss. Notice 
that mass loss is inevitable and the definition of ${\bar A}$ is meaningful 
only if $dM \neq 0$. Our sample CABs are all detached systems. Mass transfer must 
not be occurring. Nevertheless, mass loss too may produce mass ratio change, which 
we prefer to ignore for simplicity in the first approximation. Then, the orbital 
period evolution of detached CABs primarily depends on mass loss rate and the 
value of ${\bar A}$. Consequently, OAM evolution because of mass loss, needs to 
be understood first.

\subsection{OAM evolution for detached field CABs}

A more direct way of understanding OAM ($J$), orbital period and total 
mass evolution of CABs is provided by the kinematical ages of the sub-groups 
of the sample binaries. Although, the initial orbital periods and the initial 
masses are not known, the present orbital periods are confirmed to be shifted 
towards shorter periods by the increased ages of the subgroups according to 
orbital periods (see Fig. 7 of Karata\c{s} et al.\ 2004). In order to 
investigate the age dependence of the OAM ($J$), the orbital period ($P$) and 
the systemic total mass $M$ here, we have formed different sub-groups according 
to $J$ among the field detached CABs. Those sub-groups according to $J$ are 
separated by blank rows in Table 1. The mean orbital periods and the mean total 
mass of binaries in those sub-groups of the 62 field CABs are listed in Table 3. 
N in column 4 indicates the number of systems in each sub-group. 

The kinematical ages of those sub-groups were determined from their space 
velocity dispersions, using the kinematical tables of Wielen\ (1982). 
The dispersions and the implied kinematical ages are listed in the last 
two columns of Table 3, together with their estimated standard errors. 
The mean $\log J$ and the kinematical ages of the sub-groups are plotted 
in Fig. 1. In the first approximation, a linear trend line was fitted
using the least squares method. A linear trend line with a negative 
inclination in a logarithmic scale indicates an exponential decrease 
of OAM according to age. This is an observational evidence of OAM evolution 
provided by kinematical data.

The MG group binaries (Table 2), all have ages less than 0.6 Gyr, would have been 
shown by a vertical line in Fig. 1 for comparison to the field systems. 
Notice that the age is known for each star in a MG according to its moving group
membership. Instead of expressing them by a vertical line, which would 
indicate no observable evolution, we have preferred to combine the field 
and MG systems, that is, re-arrange the total sample into six sub-groups 
according to their OAM ($J$). In this way we expected to see the effect of
mixing young stars erroneously into the field systems. In other words: 
what would happen if MG systems were not selected out of a sample?  

\begin{figure}
\center
\resizebox{8cm}{8cm}{\includegraphics*{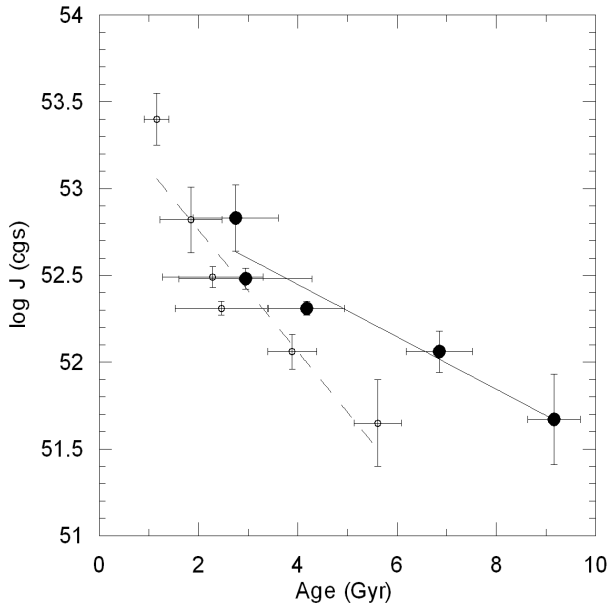}}
\caption{Orbital angular momentum $J$ versus age for the 62 
field CABs (solid line) and whole sample (dashed line) of 114 stars.}
\end{figure}

\begin{table}
\caption{The sub-groups of the 62 field CABs according to OAM.}
{\scriptsize
\begin{tabular}{cccccc}
\hline
$\log J$ & $<\log P>$ & $<\log M>$  & N & $\sigma_{tot}$  & Age \\
(cgs) &(days) & $(M_{\odot})$ & & (km $s^{-1}$)& (Gyr) \\
\hline
(50.90, 51.90]& 0.337 & 0.055 &15 & 75.49~~~3.14 & 9.16~~~0.53\\
(51.90, 52.25]& 0.472 & 0.259 &10 & 62.24~~~3.75 & 6.85~~~0.67\\
(52.25, 52.40]& 0.901 & 0.331 &11 & 47.65~~ 4.21 & 4.18~~~0.77\\
(52.40, 52.60]& 1.275 & 0.364 &14 & 40.56~~~8.16 & 2.95~~~1.34\\
(52.60, 53.25]& 1.495 & 0.509 &12 & 39.40~~~5.00 & 2.75~~~0.86\\
\hline
\end{tabular}
}  
\end{table}  

The kinematical ages of the six sub-groups of the total sample (MG+Field) have 
been re-computed from their space velocity dispersions as was done for the field 
stars. The assigned ages of MG systems in each group is ignored in this process. 
The kinematical ages and mean OAMs of these newly formed sub-groups are shown for 
comparison in Fig. 1 together with  a line (dashed) fitted using the least squares 
method. The steeper inclination of the dashed line in Fig. 1 indicates there is 
faster OAM evolution among the total sample (114) in comparison to the evolution 
(solid line) within the field CABs. Of course, faster evolution is just an illusion 
because each sub-group of the total sample contains young stars with small space 
velocities with respect to LSR (Local Standard of Rest). Therefore, the mean 
dispersions of the total sample sub-groups are reduced. Smaller dispersions, on the 
other hand, correspond to smaller kinematical ages. With smaller ages, the higher 
inclination were produced. Accordingly, it can be concluded that the rate of decrease 
of OAM would be overestimated if young MG group stars were left in the sub-groups. 
Therefore, we conclude that the sub-groups in Table 3 are corrected for this error. 
As a result, the solid line in Fig. 1 shows the corrected OAM evolution among the CABs.

Nevertheless, this correction associated with removing possible 
MG members from a sample is still a first order correction. This is because
the kinematical criteria for determining the moving group members do not 
constitute proof of membership since there is always a possibility that 
members and non-members could share the same velocity space. 
Karata\c{s} et al.\ (2004) believed non-members are negligible in number 
and do not spoil the statistics. Therefore, further purification 
by selecting out non members from the possible MG members, which requires 
independent proof of a different age or chemical composition, was not 
attempted. Since it is possible that some small number of field stars 
were erroneously selected as MG members, the inclination of the solid 
line in Fig. 1 can be considered a lower limit. This is because, 
if erroneously selected MG stars with smaller space velocity were 
put back into field stars, the ages of the sub-groups would have been
lowered accordingly. Because this second order correction was not applied,
we can only estimate that the OAM evolution among the field CABs could be
little faster, but not slower than the evolution implied in Fig. 1.

\subsection{Orbital period evolution and mass loss for detached field CABs}

The observed OAM decrease among the field CABs requires mass loss to 
carry OAM out of the systems in order to have a reducing effect 
on the orbital periods unless there 
are mechanisms which do not require mass loss. One possibility is the direct 
loss of binary binding energy by stellar encounters in the galactic space 
which is expected to be effective down to few days period (Stepien \ 1995; 
Ghez, Neugebauer \&  Matthews\ 1993). Likely it is negligible, but the other 
possibility is friction between the binary components and circum binary material 
in Keplerian orbits. It is not the scope of this study to solve which mechanism 
is dominant and what are the contributions to the orbital period evolution. 
Indeed, the statistics in Table 3 clearly indicate that total masses and orbital
periods also decrease with stellar kinematical ages similar to OAM evolution. 
The age dependence of the orbital periods $(P)$ and mean total masses 
$(M= M_{1} + M_{2})$ of the field CABs are plotted in Fig. 2, where the age 
dependence of $J$ is also shown for a comparison.

\begin{figure}
\center
\resizebox{6cm}{15.83cm}{\includegraphics*{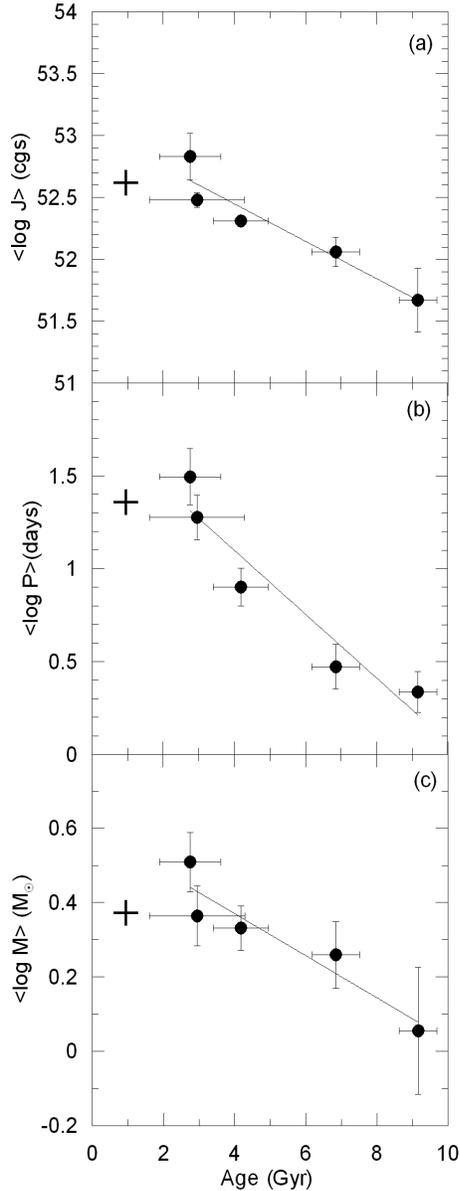}}
\caption{The age dependent variations of OAM $J$ (a), period $P$ (b) and mass 
$M$ (c) for the sub-groups of the 62 field CABs. Dot and plus show field and 
MG CABs, respectively.}
\end{figure}

The linear trend lines fitted by least squares indicate the first order 
approximations to describe the changes at $P$, $M$ and $J$. With similar arguments 
stated for $J$, the decreasing rates of systemic masses ($M$) and orbital periods 
($P$) could also be considered as lower limits. 

With ages less than 0.6 Gyr, the MG group CABs are plotted as a youngest group 
in Fig. 2. It is possible to claim that their position does not support 
the general trend of the fitted lines reasonably as their position appears lower 
than the position expected. This again could be explained by the pollution of a 
limited number of older stars in MG. The true members of MG would have agreed 
the general trends. 

Orbital period data in Fig. 2b indicate that the orbital period changes 
$(dP/dt)$ at the younger ages are faster. Such a trend, however, 
will not be consistent with the orbital period shrinkage being 
slow at the beginning and becoming fast later as described by 
Guinan \& Bradstreet\ (1988), van't Veer \&  Maceroni\ (1988, 1989), 
Stepien\ (1995) and Demircan (1999) who predicted it from the magnetic 
braking of tidally locked close binaries. Nevertheless, the appearance of 
our data could be a result of small-number statistics so that we are 
content to assume the inclination is monotonic to the first approximation.

\subsection{Decreasing rates for $J$, $P$ and $M$}

Regression analysis of linear line fitting to the $J$, $P$ and $M$ 
data (Table 3) by least squares gives the following:

\begin{eqnarray}
\log J= - (1.51 \pm 0.27) \times10^{-10} t + 53.05 \pm 0.13
\end{eqnarray}                                                                             
\begin{eqnarray}
\log P= - (1.72 \pm 0.30)\times10^{-10} t + 1.79 \pm 0.18
\end{eqnarray}    
\begin{eqnarray}
\log M= - (5.65 \pm 1.18)\times10^{-11} t + 0.60 \pm 0.07.
\end{eqnarray}    
Taking the time derivatives, 
\begin{eqnarray}
dJ/dt= -3.48\times10^{-10} J
\end{eqnarray}                                                                             
\begin{eqnarray}
dP/dt= -3.96\times10^{-10} P
\end{eqnarray}           
\begin{eqnarray}
dM/dt= -1.30\times10^{-10} M
\end{eqnarray}     
are obtained for the detached field systems, where $J$ is in cgs, $t$ in yrs, 
$P$ in days and $M$ in $M_{\odot}$. As a natural consequence of a logarithmic 
scale, the derivatives are proportional to the varying function itself. Defining 
the decreasing rate coefficients as $\alpha_{X} = -{dx \overwithdelims()dt}
{1\overwithdelims()X}$, where X could be $J$, $P$ or $M$, the rate coefficients 
for the orbital angular momentum, for the orbital period and for the total mass will 
be set respectively as $\alpha_{J}=(3.48\pm0.62)\times10^{-10}yr^{-1}$, 
$\alpha_{P}=(3.96\pm0.69)\times10^{-10}yr^{-1}$ and  $\alpha_{M}=(1.30\pm0.27)\times 
10^{-10}yr^{-1}$. Then, $J$, $P$ and $M$ can be expressed by exponential 
functions as the following after the re-integration of (4), (5) and (6):

\begin{eqnarray}
J= J_{0}~e^{-\alpha_{J} t}
\end{eqnarray}     
\begin{eqnarray}
P= P_{0}~e^{-\alpha_{P} t}
\end{eqnarray}     
\begin{eqnarray}
M= M_{0}~e^{-\alpha_{M} t},
\end{eqnarray}     
where the integration constants are evaluated as $J_{0}$, $P_{0}$ and $M_{0}$ 
which may represent OAM, orbital period and total mass of a binary at the time 
$t =0$. If $t$ represents the age of a system, then  $J_{0}$, $P_{0}$ and $M_{0}$ 
become initial OAM, orbital period and systemic total mass. If $t=0$ is taken to 
be the present date, then the future dynamical evolution of $J$, $P$ and $M$ 
could be predicted. The dynamical evolution of the 62 field CABs for the initial 
periods from 1 to 5 days has been plotted in Fig. 3.

\begin{figure}
\center
\resizebox{8cm}{5.45cm}{\includegraphics*{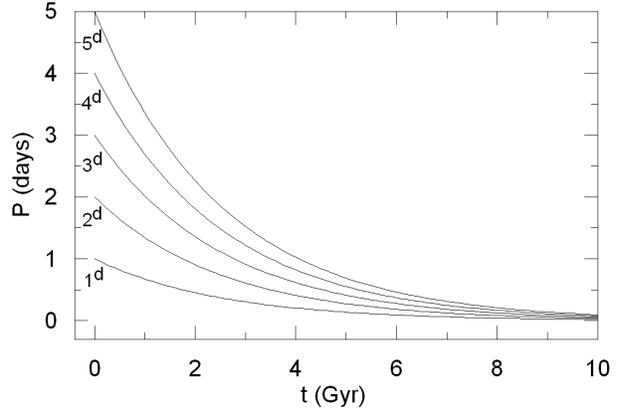}}
\caption{Dynamical evolution of the 62 field CABs for the initial periods from 
1 to 5 days.}
\end{figure}

It is advantageous to have rate coefficients as constants. So, the halving times 
could be determined as $T_{J} = 0.693/ \alpha _{J}= 1.99 \pm 0.36$ Gyr for the OAMs,  
$T_{P} = 0.693/\alpha _{P} = 1.75 \pm 0.32 $ Gyr for the orbital periods and 
$T_{M} = 0.693/\alpha _{M} = 5.33 \pm 1.16$ Gyr for the systemic masses with a 
concept similar to half lives of the radioactive elements. Then, $J$, $P$ and $M$ 
could be expressed by the half times as   

\begin{eqnarray}
X= X_{0}~e^{-{0.693 \over T_{X}}t},
\end{eqnarray}   
where $X$ could be $J$, $P$ and $M$.

Guinan \& Bradstreet\ (1988) estimated that a detached system with 
$M = 1.10 + 0.74~M_{\odot}$ and initial periods $P_{i}$ = 3.4, 2.3 and 1.2 days, 
becomes a contact binary with an orbital  period $P_{c}$ = 0.315 days at 
about 4, 1 and 0.7  Gyrs time respectively by the process of AML due 
to the magnetic braking. However, according to the period decreasing 
rate derived in this study, a similar system reaches contact configuration 
of a same contact period at about 6.01, 5.02 and 3.38 Gyrs respectively.
Except for the longest period case, the timescale differences are 
substantial. Such differences are expected since our rate was  
found statistically and it stands for an average representing 
all the orbital periods from 0.479 days to several tens of days. 
Moreover, our rates are lower limits, so our time 
scale predictions could be only upper limits.

In our estimate we found a system with $M = 1.10 + 0.74~M_{\odot}$ would lose 
54.2 per cent, 47.9 per cent and 35.6 per cent of its mass in the duration of 
6.01, 5.02 and 3.38 Gyrs of pre-contact time. Again, a substantial amount of mass 
loss seems inevitable. Even if radius change during main-sequence lifetime ($\sim 
10$ Gyr for a solar mass star) could be ignored, a better modeling with a nuclear 
and radius evolution could be prompted if one considers such an amount of mass loss. 
However, the dynamical evolutions of short-period systems like ER Vul and XY UMa 
which have components on the main-sequence, could be predicted rather consistently 
by assuming their present radii stay unchanged. 

The present orbital period of ER Vul is $P_{0}$ = 0.698 days. It is a detached 
system with two main-sequence $(G0V+G5V)$ stars having masses 
$M = 1.10 + 1.05~M_{\odot}$ and radii $R_{1} + R_{2}= 1.11 + 1.08~R_{\odot}$. 
 
The semi major axis of ER Vul at the contact configuration can be 
predicted as $a = a_{1} + a_{2} = 2.92~R_{\odot}$ by considering that 
$R_{1} + R_{2} \simeq 0.74a$ in contact configuration 
(see Kopal\ 1978). Again, with the present day 
systemic mass $M = 2.15~M_{\odot}$ and $a =  2.92~R_{\odot}$,  Kepler's third law 
would imply $P_{C}$ = 0.394 days for the contact orbital period. Then, using the 
1.75 Gyr of halving time for orbital period, eq. (22) predicts 1.44 Gyr for 
ER Vul to reach the contact configuration.

However, $\Delta T$ = 1.44 Gyr of time duration causes ER Vul to lose 17.1 
per cent of its mass. Since this mass loss has not been considered in the first 
order estimation of the $P_{C}$, the $P_{C}$ must be under and $\Delta T$ must be 
over estimated. After only five steps of iterations, $P_{C}$ = 0.427 days and 
14.9 per cent mass loss becomes consistent with  $\Delta T$ = 1.238 Gyr for 
ER Vul to reach a contact configuration. At least we could claim this is the 
upper limit according to the empirical estimate of the decreasing rates of 
$J$, $P$ and $M$ from the kinematics of CABs.    
   
\section{conclusions}

The well known spin-down of single stars due to AML requires orbital 
shrinkage (period decrease) in close binary systems. For this process to be 
effective, the spin-orbit coupling $(P_{s} \cong P_{orb})$ consequence of 
tidal locking was already suggested to be a necessary condition 
(Guinan \& Bradstreet\ 1988; van't Veer \& Maceroni\ 1988, 1989; Stepien\ 1995). 
Thus, long period binaries with no tidal locking are not expected to evolve 
into shorter period systems since tidal locking would be ineffective in 
transferring orbital angular momentum to the spinning components where 
magnetic braking operates. 

With a shrinking orbit, the more massive component of a short period binary may 
fill its Roche lobe faster and start transferring mass to the other component, 
while the system is still evolving towards shorter periods under the wind-driven 
mass loss and spin-orbit coupling mechanism. Only after Roche lobe overflow starts, 
AM evolution of the binary becomes dominated by the mass transfer. Until the 
mass-ratio reversal, the orbital period should be decreasing. But during the second 
stage, after the mass ratio reversal, the orbital period is expected to increase, 
that is, opposite to shrinking, and the orbit starts to enlarge. 

The sample of this study contains detached CABs with orbital period greater 
than 0.479 days (XY UMa). A direct period limit to tidal locking is not known. 
However, to obtain undistorted statistical results, lower limits are applied 
(2.4 days by van't Veer \& Maceroni\ 1988, and 10 days by Demircan\ 1999). 
This limit appears to be the function of the total mass, the orbital period, 
as well as age of the system (see, Tassoul\ 2000). By comparing orbital and 
rotation periods in Table 1 and Table 2, we estimated that it is not less then 
about $\sim 70$ days in the field CABs and around 10 days in the MG CABs. However, 
the MG CABs are not fully effective to give age dependent variations in $J$, $M$ and 
$P$ and there are only two systems with $P>70$ days in the field CABs (see Table 2).
Thus, we just ignored the tidal locking limit to the orbital period of field CABs.

Nevertheless, the tidal or the magnetic locking, in principle, does not involve a mass
loss directly. But, it is a mechanism which transfers OAM to the components, where 
the AM is lost by magnetically driven stellar winds. The magnetic field lines, 
especially the ones which are perpendicular to the stars surface and could reach up 
to the Alfven radius, enforces plasma to co-rotate. As long as the mass is lost
from the Alfven radius, not from the surface directly as in the case of isotropic
winds, angular momentum loss per particle appears to be amplified. Since 
OAM loss becomes associated with the mass loss and there is a mechanism which 
amplifies angular momentum loss per particle, the condition derived from 
eq. (12) will be valid.   

The amplification factor could differ from one system to another. 
However the general trend, as implied by the plots in Fig. 2, indicates 
that the average amplification ($\bar A$) must be bigger than 5/3 for 
the majority (perhaps all) of CABs so that the decrease of $P$ together 
with the decrease of $J$ and $M$ by age is observed. Otherwise, if $\bar A<5/3$, 
the orbital periods would have increased despite the mass and OAM loss.  
The average amplification factor for the present sample could be determined as 

\begin{eqnarray}
{\bar A} = {{dJ \over dt}{1\over J} \over {dM \over dt}{1\over M}} 
= {{\alpha _{J} \over \alpha _{M}} = 2.68}.
\end{eqnarray}

If such an amplification mechanism operates among the CABs, one must expect 
the rate coefficients $\alpha _{J}$, $\alpha _{M}$ and  $\alpha _{P}$ must hold a relation 
 
\begin{eqnarray}
3{\alpha _{J}} = 5{\alpha _{M}} + {\alpha _{P}}
\end{eqnarray}
according to eq. (6). Replacing $\alpha _{J}$ by $\bar A \alpha _{M}$ according 
to eq. (23), eq. (24) reduces to

\begin{eqnarray}
{(3\bar A -5)}{\alpha _{M}}= {\alpha _{P}},
\end{eqnarray}
which implies that the decreasing rate of orbital periods can also be predicted from 
the amplification parameter and the mass loss rate if they are known. 

The mean amplification factor $\bar A$ = 2.68, and the mean mass loss rate 
$\alpha_{M}=1.30\times 10^{-10}yr^{-1}$ for the present sample of CABs require 
the mean rate of orbital period decrease to be $3.95 \times 10^{-10}yr^{-1}$ 
according to above relation. It is indeed interesting that this computed value 
agrees with the value $3.96 \times 10^{-10}yr^{-1}$ which was found independently 
from the regression analysis of linear line fitting to $P$ data in Table 3. So 
we can conclude that the present data confirm the orbital period decrease as a 
cause of the mass loss from CAB systems with a mechanism which is sufficient to 
draw 2.68 times more OAM than isotropic mass loss from the component surfaces. 

As for the direct confirmation of the predicted continuous orbital period decreases by O--C 
diagrams, it is known in general that the changes as small as about $10^{-9}$ days 
in systems with periods around one day would be observable over a timespan of a 
century. However, in the case of CABs, because of the large scatter and fluctuations 
due to magnetic activity or due to third body with light-time effect in the O--C 
diagrams, it may not possible to detect our prediction of continuous orbital period decreases 
which operate in a much larger time-scale which is comparable to main-sequence 
evolutionary times.

\section{Acknowledgments}
We would like to thank Edwin Budding and anonymous referee for their useful comments.

\bsp
\end{document}